\documentclass[preprintnumbers, floatfix,letterpaper,aps,prd,epsfig,nofootinbib,
twocolumn
]{revtex4-1}
\usepackage{bm,graphicx,dcolumn,epstopdf,epsf, latexsym,mathbbol, amssymb,amsmath,color,slashed, mathrsfs,mathcomp,simplewick}
\usepackage[center]{subfigure}
\usepackage{multirow}
\usepackage{makecell}
\usepackage[colorlinks,linkcolor=blue,citecolor=blue,urlcolor=blue]{hyperref}

\begin{document}
\allowdisplaybreaks


\title{Accretion disc around black hole in Einstein-$SU(N)$ non-linear sigma model}

\author{G. Abbas $^1$ \footnote{ghulamabbas@iub.edu.pk }, Hamza Rehman  $^1$ \footnote{hamzarehman244@gmail.com}, M. Usama $^1$ \footnote{mu97947@gmail.com}, Tao Zhu $^{2, 3}$ \footnote{zhut05@zjut.edu.cn}}
\address{ ${}^1$ Department of Mathematics, The Islamia University of Bahawalpur, Bahawalpur Pakistan \\
${}^2$ Institute for Theoretical Physics \& Cosmology, Zhejiang University of Technology, Hangzhou, 310023, China\\
${}^{3}$ United Center for Gravitational Wave Physics (UCGWP),  Zhejiang University of Technology, Hangzhou, 310023, China}

\date{\today}

\begin{abstract}

The accretion of matter onto celestial bodies like black holes and neutron stars is a natural phenomenon that releases up to $40\%$ of the matter's rest-mass energy, which is considered a source of radiation. In active galactic nuclei and X-ray binaries, huge luminosities are observed as a result of accretion. Using isothermal fluid, we examine the accretion and geodesic motion of particles in the vicinity of a spherically symmetric black hole spacetime in the Einstein-$SU(N)$ non-linear sigma model. In the accretion process, the disk-like structure is produced by the geodesic motion of particles near the black hole. We determine the innermost stable circular orbit, energy flux, radiation temperature, and radioactive efficiency numerically. In the equatorial plane, we investigate the mobility of particles with stabilities that form circular orbits. We examine perturbations of a test particle by using restoring forces and particle oscillations in the vicinity of the black hole. We analyze the maximum accretion rate and critical flow of the fluid. Our findings demonstrate how parameter $N$ influences the circular motion of a test particle as well as the maximum accretion rate of the black hole in the Einstein-$SU(N)$ non-linear sigma model.

\end{abstract}

\maketitle
\section{Introduction}

The theory of general relativity (GR) predicts the existence of black holes (BHs), which are remarkably, extremely massive objects. These are the source of the strongest gravitational field in the universe and have magnetic intensity and high spin. Because of these properties, BHs are an ideal laboratory for astrophysical studies of matter and gravity. Some observational data has confirmed the existence of BH in recent years. The first was the discovery of gravitational waves from a binary BH merger by the collaboration of LIGO and Virgo \cite{a1}. A further important contribution is the Event Horizon Telescope's use of very long Baseline Interferometry to get the first images of M87 BH shadow \cite{a2, a3} and the recently released image of Sgr $A^{*}$ \cite{a4}. These significant accomplishments help us to understand GR theory and the nature of accretion discs around supermassive celestial bodies in the strong gravity regimes around the event horizon.

It is speculated that celestial bodies, such as BHs, accumulate mass through the accretion process. These might be used to test modified theories of gravity. The presence of an accretion disc is also a necessary component to maintain the high accretion rate. By slowly spiraling into a centrally compact object, diffuse material forms an accretion disc and radiates gravitational energy. The phenomenon by which a dense object like a BH gathers particles from a nearby fluid is called accretion. These particles must pass through a critical point when they begin to accelerate, where the fluid velocity equals the sound speed. At supersonic speeds, the fluid flows onto the central mass. As a result of this, the BH mass should be increasing \cite{a5}. Also, this phenomenon released energy, which could be the source of some astronomical events like the formation of jets, quasars, and radiation \cite{a6}. As a result of analyzing the geodesic structure of particles in the vicinity of the BH, it is interesting to analyze several typical radii, such as marginally bound orbit ($r_{\rm mb}$) and innermost stable circular orbit (ISCO). These radii are significant in the investigation of BH accretion discs.

In accretion discs, the interior edge of the disc corresponds to ISCO, and the efficiency of the energy emitted is a representation of how well rest-mass energy is converted into radiative energy \cite{a7}, which may be calculated using these radii. The maximum or minimum of the effective potential is associated with the positions of unstable or stable circular orbits, respectively. According to Newtonian theory, the ISCO can have an arbitrary radius after the effective potential reaches a minimum value with any value of the angular momentum, indicating that there is no minimum radius for the ISCO \cite{a8}. However, the situation differs once the effective potential demonstrates general relativistic effects or has a significant feature that depends on the angular momentum of particles and other parameters. For instance, the effective potential in GR and for particles revolving around the Schwarzschild BH has two extremal values for any value of the angular momentum (minimum or maximum). Consequently, the two points exclusively correspond to a particular value of angular momentum. This point examines ISCO at $r = 3r_{g}$ \cite{a8, a9}, whereas Schwarzschild radius is represented by $r_{g}$.

The characteristics of spacetime affect the positions of these radii in distinct metrics, and several quantities such as angular velocity, angular momentum, and specific energy are significant in the location of these points. Because of this, the study of the particle dynamics and accretion process near black holes with different black hole spacetimes has attracted a lot of attention. Ruffini et al.\cite{b1} and Bardeen et al. \cite{b2} have investigated the impacts of ISCO near the Kerr BH. In their textbook on GR, Hobson et al. \cite{b3} have established these characteristics. The efficiency of Kerr and Schwarzschild BH accretion discs was calculated by Thorne and Novikov \cite{b4}. Psaltis and Johannsen \cite{b5} proposed the Kerr-like metric, and Johannsen \cite{b6} developed the accretion discs near such BHs. Tursunov et al. \cite{b7} discovered the geodesic structure and circular orbits of charged particles around rotating BHs that are weakly magnetized. Because particles in an accretion disc rotate in stable orbits, whenever perturbations affect the particles, oscillations are obtained in vertical and radial directions with epicyclic frequencies. As a result, the analysis of orbital and epicyclic frequencies is essential to the physics of the accretion discs surrounding BHs. Isper \cite{b8,b9}, Wagoner \cite{c1}, Kato \cite{c2}, and Ortega-Rodriges et al. \cite{c3} have conducted investigations in this area. Tretyakova \cite{c4} has discussed geodesics for scalar-tensor empirical characteristics in Horndeski BH. Additionally, Salahshoor and Nozari \cite{c5} studied the circular orbits and accretion discs in a class of Horndeski/Galileon BHs using a similar approach. Moreover, in \cite{c6}, Abbas and Ditta have investigated the circular accretion near a regular phantom BH. Recently, the particle dynamics and thin accretion disk around black holes in the Einstein-\AE{}ther theory have also been explored \cite{Azreg-Ainou:2020bfl, Liu:2021yev, He:2022lrc, Wang:2022yvi}.

In this paper, we investigate the accretion disc of a black hole space-time in the Einstein-$SU(N)$ non-linear sigma model (NLSM). The non-linear sigma model is one of the most important theoretical frameworks with applications in many contexts, for example in the dynamics of Goldstone bosons \cite{bosons}, in condensed matter systems \cite{string}, in string theory \cite{string}, and in supergravity \cite{supergravity}. One relevant example is the $SU(N)$-NLSM with $N=2$, which provides an effective description of the low-energy dynamics of pions in $3+1$ dimensions \cite{pions}. Thus, it is interesting to explore the coupling of the $SU(N)$-NLSM to GR both from the theoretical and the phenomenological point of view.  With this motivation, the Einstein-NLSM system has been studied in a lot of works \cite{Canfora:2013osa, Giacomini:2017xno, Astorino:2017rde, Astorino:2018dtr}. Recently, a black hole solution is obtained in Einstein-$SU(N)$-NLSM theory \cite{Henriquez-Baez:2022ubu}.  The effects of the coupling constant of $SU(N)$-NLSM theory on the thermodynamics and stability of this black hole have also been explored \cite{Henriquez-Baez:2022ubu}. It is natural to ask whether the $SU(N)$-NLSM model can have any effects on particle dynamics and the accretion process near the black hole, and this is exactly the main purpose of the current paper.

To study the particle dynamics of the black hole in Einstein-$SU(N)$-NLSM theory, we limit our investigation to the polar coordinates system and equatorial plane for the conserved quantities of the test particles. Furthermore, we determine the event horizon and singularity of the BH. Moreover, we study circular orbits and effective potential to investigate the locations of circular orbits such as the photon sphere ($r_{ph}$), the $r_{mb}$, and $r_{isco}$. Ultimately, the critical accretion is determined by using the isothermal fluid parameters. This paper is arranged as follows. In section \textbf{2}, we provide a brief introduction of the Einstein-$SU(N)$-NLSM theory and its black hole solution obtained in \cite{Henriquez-Baez:2022ubu}. In section \textbf{3}, the general formulation for the mobility of particles is discussed, where circular motion, radiant energy flux, stable circular orbits, and oscillations are explored. The general formulas for several dynamical parameters, including accretion rate, critical flow speed, and accretion for an isothermal fluid, are determined in section \textbf{4}. In section \textbf{5}, we analyzed the solution of the black hole in the Einstein-$SU(N)$-NLSM and a circular geodesic in the equatorial plane. Finally, section 6 addresses conclusions and discussions.

\section{Black hole spacetime in Einstein-$SU(N)$ nonlinear sigma model}

In this section, we present a brief introduction to the black hole in Einstein-$SU(N)$-NLSM. The action of the Einstein-$SU(N)$-NLSM is given by \cite{Henriquez-Baez:2022ubu}
\begin{equation}
I[g,U]=\int d^{4}x\sqrt{-g}\left(\frac{R}{2\kappa}+\frac{K}{4}Tr[L^{\mu}L_{\mu}]\right),\label{za1}
\end{equation}
where $R$ is the Ricci scalar and $L^{\mu}$ is Maurer-Cartan form components given by
\begin{equation}
L_{\mu}=U^{-1}\partial_{\mu}U=L_{\mu}^{i}t_{i},\label{za2}
\end{equation}
with $U(x)\in SU(N)$. Here, $N$ denotes the flavor number, and $t_{i}$ represents the $SU(N)$ Lie group generators, where $i = 1$,... $(N^{2}-1)$. Hence, the gravitational constant is represented by $\kappa$, and $K$ is an experimentally determined positive coupling constant. Varying the action in Eq. (\ref{za1}) with respect to the fields $U$ and $g_{\mu\nu}$ yields the field equations,
\begin{equation}
\nabla^{\mu}L_{\mu}=0,\label{za3}
\end{equation}
\begin{equation}
G_{\mu\nu}=\kappa T_{\mu\nu},\label{za4}
\end{equation}
where $\nabla^{\mu}$ is Levi-Civita covariant derivative, $G_{\mu\nu}=R_{\mu\nu- \frac{1}{2}g_{\mu\nu}R}$ is the Einstein tensor, and $T_{\mu\nu}$ is the energy-momentum tensor of the $SU(N)$-NLSM which is given by
\begin{equation}
T_{\mu\nu}=-\frac{K}{2}Tr[L_{\mu}L_{\nu}-\frac{1}{2}g_{\mu\nu}L^{\alpha}L_{\alpha}]. \label{za5}
\end{equation}
Note that Eq. (\ref{za3}) contains $(N^{2} - 1)$ non-linear coupled second order differential equations.

The spherically symmetric BH spacetime was derived in \cite{Henriquez-Baez:2022ubu} and the metric is given as
\begin{equation}
ds^{2}=f(r)dt^{2}-\frac{1}{f(r)}dr^{2}-r^{2}d\theta^{2}-r^{2}\sin^{2}\theta d\phi^{2},\label{ya4}
\end{equation}
with
\begin{equation}
f(r)=1-K \kappa a_{N}-\frac{2m}{r}-\frac{\Lambda}{3}r^{2}, \label{ya6}
\end{equation}
where $m$ denotes the constant of integration and $a_{N}$ is given by
\begin{equation}
a_{N}=\frac{N(N^{2}-1)}{6}.\label{ya3}
\end{equation}
The given configuration clearly illustrates by pionic matter and generalization of $SU(N)$ BH obtained in Refs. \cite{d7} and \cite{d8}. Furthermore, this solution is an asymptotically anti-de Sitter (AdS) form of the Barriola-Vilenkin metric \cite{d9}, and one can recover Schwarzschild-AdS BH by setting $K = 0$.

{It is interesting to note that this solution also reduces to the BH solution in the Einstein-$SU(2)$-Skyrme theory with $N=2$ and without the Skyrme coupling. When the Skyrme coupling is present, it gives an explicit contribution to the metric function $f(r)$ of order $1/r^2$. The effects of such Skyrme coupling on the deflection angle of light around the BH in the Einstein-$SU(2)$-Skyrme theory has been studied in \cite{Canfora:2018isz}. Thus it is natural to seek how the above solution can be modified by including a Skyrme coupling in the Einstein-SU(N)-NLSM model. As mentioned in \cite{Henriquez-Baez:2022ubu} and using a similar formalism, the above solution can be generalized to solutions of the SU(N)-Skyrme model and an analytical form of it has not been reported yet. Since the main purpose of the current paper is not intended to construct a solution with the Skyrme coupling, we will only focus on the above solution without the contribution from the Skyme term.}

The event horizon of the BH is located at
\begin{widetext}
\begin{eqnarray}
r_{+}=\frac{-\Lambda+K\kappa a_{N}\Lambda-(3m \Lambda^{2}+\sqrt{\Lambda^{3}((K\kappa a_{N}-1)^{3}+9m^{2}\Lambda)})^{\frac{2}{3}}}{\Lambda(3m \Lambda^{2}+\sqrt{\Lambda^{3}((K\kappa a_{N}-1)^{3}+9m^{2}\Lambda)})^{\frac{1}{3}}}.
\end{eqnarray}
\end{widetext}
It is observed that, in order to get a real root in the earlier result, the constant of integration $m$ must fulfill the relation
\begin{eqnarray}
m\geq\frac{({K\kappa a_{N}-1)}^{\frac{3}{2}}}{3\sqrt{-\Lambda}}.
\end{eqnarray}
The above relationship establishes the event horizon's minimum radius, namely
\begin{eqnarray}
r_{min}=\frac{2\sqrt{K\kappa a_{N}-1}}{\sqrt{-\Lambda}}.
\end{eqnarray}
The horizon radius is depicted in $\textbf{Fig.1}$ for distinct values of $N$.

\begin{figure}
\includegraphics[width=8.0cm]{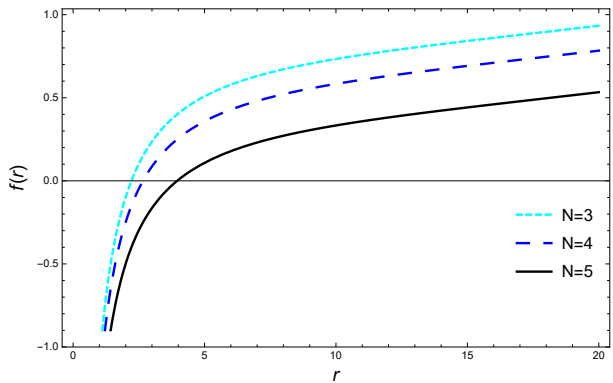}
\caption{ The plot between $f(r)$ and $r$ for distinct values of $N$}.
\end{figure}

\section{General formalism for test particle's motion}

In this portion, we derive the general formalism for test particles by considering the spherically symmetric and static space-time. We consider two killing vectors $\xi_{t}=\partial_{t}$ and $\xi_{\phi}=\partial_{\phi}$ along with two constants of motion $L$ and $E$ (angular momentum and conserved energy), which are given as follows
\begin{equation}
E=-g_{\mu\nu} \xi_{t}^{\mu} u^{\nu}\equiv -u_{t},\label{a5}
\end{equation}
and
\begin{equation}
L=g_{\mu\nu} \xi_{\phi}^{\mu}u^{\nu}\equiv u_{\phi},\label{a6}
\end{equation}
where the four-velocity for test particle is $u^{\mu}=\frac{dx^{\mu}}{d\tau}=(u^{t}, u^{r} ,u^{\theta}, u^{\phi})$ and satisfy the normalization condition $u^{\mu}u_{\mu}=1$, which can be simplified to
\begin{equation}
[g_{rr}(u^{r})^{2}+g_{\theta\theta}(u^{\theta})^{2}]=[1-g^{tt}(u_t)^{2}-g^{\phi\phi}(u_{\phi})^{2}].\label{a7}
\end{equation}
In equatorial plane (i.e. $\theta=\frac{\pi}{2}$), by using Eqs. (\ref{a5}), (\ref{a6}) and (\ref{a7}), we get
\begin{equation}
u^t=-\frac{E}{f(r)},\label{a8}
\end{equation}
\begin{equation}
u^\theta=0,\label{a9}
\end{equation}
\begin{equation}
u^\phi=-\frac{L}{r^2},\label{b1}
\end{equation}
\begin{equation}
u^r=\sqrt{-f(r)\left(1-\frac{E^2}{f(r)}+\frac{L^2}{r^2}\right)}. \label{b3}
\end{equation}
From Eq.(\ref{b3}), we have
\begin{equation}
(u^r)^2+V_{eff}=E^2,\label{b4}
\end{equation}
where $V_{eff}$ is the effective potential  for the motion of the particle, which can be expressed as
\begin{equation}
V_{eff}=f(r) \left[1+\frac{L^2}{r^2}\right].\label{b5}
\end{equation}
It is obvious that the effective potential depends on spacetime metric function $f(r)$, angular momentum of particles, and radial velocity. The evaluation of effective potential plays an important role in the geodesic motion of particles because the location of ISCO is obtained by local minima of the effective potential.

\subsection{Circular Motion of test particles}

In this subsection, we examine the circular motion of particles in the equatorial plane, so $u^{r}=\dot{u}^{r}=0$ must be fulfilled because the radial component $r$ is necessarily constant. From Eq. (\ref{b4}), we have $V_{eff}=E^2$ and also $\frac{d}{dr}V_{eff}=0$. We figure out angular velocity $\Omega_{\phi}$, specific energy $E$, specific angular momentum $L$ and angular momentum $l$, which are given by
\begin{eqnarray}
\Omega_{\phi}^2=\frac{1}{2r}f'(r),\label{b6}
\end{eqnarray}
\begin{eqnarray}
E^2=\frac{2f^2(r)}{2f(r)-r f'(r)},\label{b7}
\end{eqnarray}
\begin{eqnarray}
L^2=\frac{r^3f'(r)}{2f(r)-r f'(r)},\label{b8}
\end{eqnarray}
\begin{eqnarray}
l^2=\frac{r^3f'(r)}{2f^2(r)}.\label{b9}
\end{eqnarray}
From Eqs. (\ref{b7}) and (\ref{b8}), it is clear that the specific energy $E$ and specific angular moment $L$ must be real if
\begin{eqnarray}
2f(r)-r f'(r)>0. \label{c1}
\end{eqnarray}
The above expression is mandatory for the circular orbits because we can figure out the particular area of circular orbits to find the solution to inequality. For bound orbits and marginally bound orbits, Eq.~(\ref{b7}) must fulfil the conditions $E^{2}<1$ and $E^{2}=1$, respectively. From Eq. (\ref{b7}), we get
\begin{eqnarray}
r f'(r)+2f(r)[f(r)-1]=0.  \label{c2}
\end{eqnarray}
We can acquire marginally bound orbits by simplifying the above equation. From Eqs.~(\ref{b7}) and (\ref{b8}), it can be observed that specific energy $E$ and specific angular momentum $L$ diverge if the following relation is satisfied
\begin{eqnarray}
2f(r)-r f'(r)=0. \label{c3}
\end{eqnarray}
By using Eq. (\ref{c3}), we can calculate the radius of the photon sphere, which plays a vital role to examine gravitational lensing.

\subsection{Radiant Energy Flux and Circular Orbits }

The existence of stable circular orbits relies on the local minima of the effective potential that is obtained from $V_{eff}=0=\frac{d}{dr}V_{eff}$ and $\frac{d^{2} V_{eff}}{dr^{2}}>0$. By using Eq.~(\ref{b5}), we obtain
\begin{eqnarray}
\frac{d^2}{dr^2}V_{eff}=(1+\frac{L^2}{r^2})f''(r)-\frac{4L^2}{r^3}f'(r)+\frac{6L^2}{r^4}f(r).  \nonumber \\\label{c4}
\end{eqnarray}
By using the the conditions $V_{eff} =0$, $\frac{dV_{eff}}{dr} =0$ and $\frac{d^{2}V_{eff}}{dr^{2}} =0$, one can determine the radius of ISCO $r_{isco}$.
If $r<r_{isco}$, then the accretion process is feasible. When particles fall from rest to infinity and fall into a black hole, these would release gravitational energy, which can be converted into radiations by the source of the powerful astronomical events. In \cite{h2}, the energy flux radiant through the accretion disc can be represented in terms of $\Omega_{\phi}$ angular velocity, the specific energy $E$, and the specific angular momentum $L$ as follows
\begin{eqnarray}
K=-\frac{\dot{M} \Omega_{\phi},_{r}} {{4\pi\sqrt{-g} (E-L \Omega_{\phi})^2}}\int^{r}_{r_{isco}}(E-L \Omega_{\phi})L,_{r}dr. \nonumber \\ \label{c5}
\end{eqnarray}
Here, $K$ is the radiant flux, $\dot{M}$ is the mass accretion rate, $\Omega_{\phi},_{r} \equiv \frac{d\Omega_{\phi}}{dr}$ and $g$ is the determinant of $g_{\mu\nu}$ given by
\begin{eqnarray}
g=\det(g_{\mu\nu})=-r^4\sin^2\theta,  \label{c6}
\end{eqnarray}
where we put $\sin\theta=\sin\frac{\pi}{2}=1$ as we are dealing in equatorial plane. By using Eqs.~(\ref{b6}-\ref{b8}), we have
\begin{widetext}
\begin{eqnarray}
K(r)=\frac{-\dot{M}}{4\pi r^4} \sqrt{\frac{r}{2f'(r)}}\times \frac{[r f''(r)-f'(r)][r f'(r)-2f(r)]}{[2f(r)+r f'(r)]^2}  \int^{r}_{r_{isco}} F(r)dr,  \label{c7}
\end{eqnarray}
with
\begin{eqnarray}
F(r)=\sqrt{\frac{r}{2f'(r)}}\times  \frac{[2f(r)+r f'(r)][r f(r)f''(r)-2r f'^2(r)+3f(r)f'(r)]} {[2f(r)-r f'(r)]^2}.  \label{c8}
\end{eqnarray}
\end{widetext}
The associated expression $K(r)=\sigma T^{4}(r)$ contains energy flux as well as temperature because the accretion disc assumes to be in thermal equilibrium, so the radiation emitted from the disk surface is assumed as black body radiation. By assuming thermal black body radiation, it is quite easy to attain the temperature distribution on the disc using this relation, also it helps to obtain luminosity $L(\nu)$ of
the disc. The disc inclination angle $\gamma$ along with $d$ distance observed by the luminosity of the accretion disc \cite{R3} is given by
\begin{eqnarray}
L(\nu)=4\pi d^2 I(\nu)=\frac{8}{\pi}(\cos\gamma) \int^{r_{f}}_{r_{i}} \int^{2\pi}_{0} \frac{\nu_{e}^{3} r d\phi dr} {e^\frac{\nu_{e}} {T}-1}.\label{c9}
\end{eqnarray}
In the above expression, thermal energy flux is expressed by $I(\nu)$. In the mass accretion process, the efficiency of accreting material is one of the most essential factors. We can obtain the maximum efficiency, $\eta^{*}$ by the ratio of the specific binding energy of the ISCO to the specific rest mass energy illustrated given by
\begin{eqnarray}
\eta^{*}=1-E_{isco}.\label{d1}
\end{eqnarray}
where $E$ represents the specific energy of a particle revolving in ISCO. This relationship is true if all photons emitted are capable of escaping to infinity. If a fluid element is perturbed, the motion of a test particle will be in a circular orbit in the equatorial plane. In the following section, we will investigate the problem of the particles oscillating in circular orbits with three components of motion.

\subsection{Oscillations}

By virtue of restoring forces, several forms of oscillatory motion are observed in the accretion process. Horizontal and vertical oscillations are produced when restoring forces affect perturbation in the accretion discs. A few of the restoring forces in accretion discs are caused by the rotation of the disc as well as a vertical gravitational field. Due to a restoring force carried by the fluid's rotation, when a fluid element is transferred in the radial direction, it will eventually restore its equilibrium state. For the sake of central objects, the gravitational force in accretion discs counter balances the centrifugal force. The flow element would be dragged within or outside the original radius with epicyclic frequency, $\Omega_{r}$, once the former prevails compared to the latter or when the reverse happens. If the fluid element is perturbed vertically in the equatorial plane, the gravitational field squeezes the perturbed element into the equilibrium position. As a consequence of restoring force, the fluid element produces harmonic oscillations across the equatorial plane with vertical epicyclic frequency $\Omega_{\theta}$. In an accretion disc, the movement of the particles in the fluid depends on circular motion with orbital frequency, harmonic radial motion with radial frequency, and harmonic vertical motion with vertical frequency. Thus, we discuss the analysis of vertical and radial motion on nearby circular orbits in the equatorial plane. Now we elaborate the radial motion $\frac{1}{2}(\frac{dr}{dt})^2=V_{eff}^{(r)}$ and vertical motion $\frac{1}{2}(\frac{d\theta}{dt})^2=V_{eff}^{(\theta)}$ of the particles. From Eq. (\ref{a7}), we take $u^{\theta}=0$ and $u^{r}=0$ for vertical and radial motion. By taking $u^{r}=\frac{dr}{d \tau}=\frac{dr}{dt}u^{t}$ and $u^{\theta}=\frac{d\theta}{d\tau}=\frac{d\theta}{dt}u^{t}$, we have
\begin{eqnarray}
\frac{1}{2}\left(\frac{dr}{dt}\right)^{2}=-\frac{1}{2}\frac{f^3(r)}{E^2}\left[1-\frac{E^{2}}{f(r)}+\frac{L^{2}}{r^{2}\sin^{2}(\theta)}\right]=V_{eff}^{(r)}, \nonumber \\\label{d2}
\end{eqnarray}
\begin{eqnarray}
\frac{1}{2}\left(\frac{d\theta}{dt}\right)^2=-\frac{1}{2}\frac{f^2(r)}{E^2 r^2}\left[1-\frac{E^2}{f(r)}+\frac{L^2}{r^2 \sin^2\theta}\right]=V_{eff}^{(\theta)}. \nonumber \\ \label{d3}
\end{eqnarray}
In the equatorial plane, to examine the solution of vertical epicyclic and radial frequencies nearby circular orbits for small perturbations $\delta r$ and $\delta \theta$. We shall take time derivative of Eqs.(\ref{d2}) and (\ref{d3}), which yields
\begin{eqnarray}
\frac{d^2 r}{dt^2}=\frac{d V_{eff}^{(r)}}{dr}.\label{d3}
\end{eqnarray}
The equation of motion for a perturbed particle is
\begin{eqnarray}
\frac{d^2}{dt^2} (\delta r)=\frac{d^2 V_{eff}^{(r)}}{dr^2} (\delta r) \Rightarrow (\delta \ddot{r})+\Omega_{r}^2(\delta r)=0,\label{d4}
\end{eqnarray}
 where $\Omega_{r}^2\equiv -\frac{d^2}{d r^2} V_{eff}^{(r)}$, and $\delta r=r-r_{0}$  is perturbation radius that is displaced from its original radius $r=r_{0}$  and dot indicates the time derivative. Also $\delta \theta=\theta-\theta_{0}$ is the deviation for perturbation in the vertical direction, we have
\begin{eqnarray}
\frac{d^2 (\delta\theta)}{dt^2}=\frac{{d^2 V_{eff}}^{(\theta)}}{dr^2} (\delta\theta) \Rightarrow (\delta \ddot{\theta})+\Omega_{\theta}^2(\delta \theta)=0,\label{d5}
\end{eqnarray}
where $\Omega_{\theta}^2\equiv -\frac{d^2}{d\theta^2} V_{eff}^{(\theta)}$. By Eqs.(\ref{d2}) and (\ref{d3}) in equatorial plane, we have
\begin{widetext}
\begin{eqnarray}
\Omega_{r}^2&=&\frac{1}{2E^2 r^4} [((r^2+L^2)3f(r)-2 E^2 r^2)r^2 f(r)f''(r)+2r^2((r^2+L^2)3f(r)-E^2 r^2)f'^2(r) \nonumber \\
&&~~~~~~~~~~  -6L^2 f^2(r)(2rf'(r)-f(r))].\label{d6}
\end{eqnarray}
\end{widetext}
and
\begin{eqnarray}
\Omega_{\theta}^2=\frac{f^2(r)L^2}{E^2 r^4}.\label{d7}
\end{eqnarray}
where in Eq.(\ref{d6}) prime indicates derivative for radial coordinate $r$. Basic dynamical equations for the black hole in the Einstein-$SU(N)$-NLSM theory are described in the following section.

\section{Basic Dynamical Equations}

 We analyze the basic formulation of accretion, which was determined by Babichev et al. \cite{R4,R5}. We consider ideal fluid, which is described by energy-momentum tensor
\begin{eqnarray}
T^{\mu\nu}=(p+\rho)u^{\mu}u^{\nu}-pg^{\mu\nu},\label{d8}
\end{eqnarray}
where pressure and energy density of the fluid are represented by $p$ and $\rho$ also the four-velocity of the fluid elements is $u^{\mu}$. In the equatorial plane, the four-velocity is
\begin{eqnarray}
u^{\mu}=\frac{dx^\mu}{d\tau}=(u^t,u^r,0,0),\label{d9}
\end{eqnarray}
here $\tau$ is the proper time along the geodesic. By using normalization condition $(u^{\mu}u_{\mu}=1)$ and above expression, we acquire
\begin{eqnarray}
u^t=\frac{\sqrt{(u^r)^2+f(r)}}{f(r)}.\label{e1}
\end{eqnarray}
where the condition $u^{t}>0$ is for forward flow in time while for accretion (inward flow) $u^{r}<0$. For analyzing the accretion process, all the fundamental equations are determined by calculating the energy-momentum conservation equations as well as the particle-number conservation equations. Now the law of conservation for energy-momentum tensor is
\begin{eqnarray}
T_{;\mu}^{\mu \nu}=0 \Rightarrow T_{;\mu}^{\mu \nu}=\frac{1}{\sqrt{-g}}(\sqrt{-g}T^{\mu\nu})_{,\mu}+ \Gamma_{\alpha\mu}^\nu T^{\alpha\mu}=0, \nonumber \\ \label{e2}
\end{eqnarray}
where $\sqrt{-g}=r^{2} \sin\theta$, $\Gamma$ expresses Christoffel symbol's of 2nd kind and $(;)$ is the representation for covariant differentiation.
By using metric of BH, Eq.(\ref{e2}) becomes
\begin{eqnarray}
T_{,r}^{10}+\frac{1}{\sqrt{-g}} T^{10} (\sqrt{-g})_{,r}+2 \Gamma_{01}^{0}T^{10}=0,\label{e8}
\end{eqnarray}
after certain simplifications, we have
\begin{eqnarray}
\frac{d}{dr}[(p+\rho)u^r r^2 {\sqrt{(u^r)^2+f(r)}}]=0,\label{e9}
\end{eqnarray}
integrating the above equation, we obtain
\begin{eqnarray}
(p+\rho)u^r r^2 {\sqrt{(u^r)^2+f(r)}}=C_{0},\label{f1}
\end{eqnarray}
where $C_{0}$ is the constant of integration. By applying the law of conservation adjacent to the four-velocity through
$u_{\mu}T_{;\nu}^{\mu \nu}=0$, we get
\begin{eqnarray}
(p+\rho)_{,\nu} u_{\mu} {u^\mu u^\nu}+(p+\rho) u_{;\nu}^{\mu} {u_\mu u^\nu}  \nonumber \\ +(p+\rho) {u_{\mu} u^\mu} u_{;\nu}^\nu+p_{,\nu}g^{\mu\nu} u_\mu+p{u_\mu} g_{;\nu}^{\mu\nu}=0.\label{f2}
\end{eqnarray}
Since $g_{;\nu}^{\mu\nu}=0$ and by using normalization condition $(u^{\mu}u_{\mu}=1)$, we obtain
\begin{eqnarray}
(p+\rho){u_{;\nu}^\nu}+u^{\nu}{\rho,_\nu}=0,\label{f3}
\end{eqnarray}
as $A_{;a}^b=\partial_a A^b+\Gamma_{ac}^b A^c$, we get
\begin{eqnarray}
u^{r} \rho_{,r}+(p+\rho)[\Gamma_{0c}^0 u^c+u_{,r}^r+\Gamma_{1c}^{1}u^c) \nonumber \\ +\Gamma_{2c}^{2}u^c+\Gamma_{3c}^{3}u^c]=0.\label{f4}
\end{eqnarray}
After a few simplifications and considering non-zero components of Eq. (\ref{f4}), we get
\begin{eqnarray}
\frac{\rho'}{(p+\rho)}+\frac{u'}{u}+\frac{2}{r}=0,\label{f5}
\end{eqnarray}
by integrating the above equation, we have
\begin{eqnarray}
r^2 u^r \exp\Big(\int{\frac{d\rho}{p+\rho}}\Big)=-C_1. \label{f5}
\end{eqnarray}
Here $C_{1}$ is a constant of integration. Since $u^{r}<0$, so in above equation $C_{1}>0$ .
We compute finally
\begin{eqnarray}
(p+\rho)\sqrt{(u^{r})^2+f(r)}\exp(-\int{\frac{d\rho}{p+\rho}})=C_2,\label{f6}
\end{eqnarray}
where $C_{2}$ is a constant of integration and the expression for mass flux is given by
\begin{eqnarray}
(\rho u^\mu)_{;\mu}\equiv \frac{1}{\sqrt{-g}}(\sqrt{-g} \rho u^{\mu})_{,\mu}=0, \label{f7}
\end{eqnarray}
Eq.~(\ref{f7}) can be written as
\begin{eqnarray}
\frac{1}{\sqrt{-g}}(\sqrt{-g}\rho u^{\mu})_{,r}+\frac{1}{\sqrt{-g}}(\sqrt{-g}\rho u^{\theta})_{,\theta}=0.\label{f8}
\end{eqnarray}
We neglect $\frac{1}{\sqrt{-g}}(\sqrt{-g}\rho u^{\theta})_{,\theta}$ from Eq.~(\ref{f8}) because our concern is in equatorial plane.
So, $\sqrt{-g}\rho u^{\mu}$ would be treated as constant and we have
\begin{eqnarray}
\rho u^r r^2=C_3. \label{f9}
\end{eqnarray}
In the above equation $C_{3}$ is the constant of integration. Now, we have to describe the characteristics of dynamical parameters,
accretion rate, and critical accretion.
\subsection{Dynamical Parameters}
We consider isothermal fluids with $p=k\rho$ here $k$ is the equation of state parameter. For isothermal fluid, the speed of sound must be constant and $p\propto \rho$ also by using Eqs.~(\ref{f5}), (\ref{f6}) and (\ref{f9}), we can find
\begin{eqnarray}
\frac{p+\rho}{\rho} \sqrt{(u^{r})^{2}+f(r)}=C_4,\label{g1}
\end{eqnarray}
where $C_{4}$ is a constant of integration. By inserting $p=k\rho$ in Eq. (\ref{g1}), we deduce
\begin{eqnarray}
u=\left(\frac{1}{k+1}\right) \sqrt{C_{4}^{2}-f(r)(k+1)^{2}}. \label{g2}
\end{eqnarray}
Thus from Eq. (\ref{f9}), we get
\begin{eqnarray}
\rho=\frac{C_{3}}{r^{2}} \frac{(k+1)}{\sqrt{C_{4}^{2}-f(r)(k+1)^{2}}}.\label{g3}
\end{eqnarray}
Also by using $p=k\rho$, pressure $p$ can be determined.
\subsection{Mass Evolution}
In astronomical observation, the mass of the BH would progressively change due to a variety of events including Hawking radiation and accreting mass from the accretion disc onto the BH. We can determine the rate of change of mass by integrating the flux of fluid over the BH surface which is
$\dot{M}\equiv \frac{dM}{dt}=-\int T_{t}^{r}ds$ where $ds=\sqrt{-g}d\theta d\phi$ and $T_{t}^{r}=(p+\rho)u_{t}u^{r}$.
By putting these expressions, we obtain $\dot{M}$ as follows
\begin{eqnarray}
\dot{M}=-4\pi r^{2}u(p+\rho) \sqrt{u^{2}+f(r)} \equiv -4\pi C_{0}.\label{g4}
\end{eqnarray}
By considering $C_{0}=-C_{1}C_{2}$ and $C_{2}=(p_{\infty}+\rho_{\infty}) \sqrt{f(r_{\infty})}$, above relation proceeds to
\begin{eqnarray}
\dot{M}=4\pi C_{1}(p_{\infty}+\rho_{\infty}) \sqrt{f(r_{\infty})}M^{2}.\label{g5}
\end{eqnarray}
We can acquire the time evolution of the mass of the BH with initial mass $M_{i}$ and Eq. (\ref{g5}) can be revised as
\begin{eqnarray}
\frac{d M}{M^{2}}=\mathcal{F}t,\label{g6}
\end{eqnarray}
where $\mathcal{F}\equiv 4\pi C_{1}(p+\rho) \sqrt{f(r_{\infty})}$. Integration of Eq.(\ref{g6}) yields
\begin{eqnarray}
M_{t}=\frac{M_{i}}{1-\mathcal{F} M_{i}t}\equiv \frac{M_{i}}{1-\frac{t}{t_{cr}}},\label{g7}
\end{eqnarray}
where $t_{cr}=[4\pi C_{1}(p+\rho)\sqrt{f(r_{\infty})}M_{i}]^{-1}$ is the time of accretion. From Eq.(\ref{g7}), it is clear that at $t=t_{cr}$
the BH mass grows up in finite time up to infinity.
\subsection{Critical Accretion}
The fluid flow is static at a large distance from the BH but its movement is inward because of BH gravitational field. When the fluid moves inward it passes over the sonic point at this point fluid velocity must be equal to the speed of sound. By using Eqs.(\ref{f9}) and (\ref{g1}), we can get
\begin{eqnarray}
\frac{\rho'}{\rho}+\frac{u'}{u}+\frac{2}{r}=0,\label{g8}
\end{eqnarray}
and
\begin{eqnarray}
\frac{\rho'}{\rho}\left[\frac{d\ln(p+\rho)}{d\ln\rho}-1\right]+\frac{u u'}{u^{2}+f(r)}+\frac{1}{2} \frac{f'(r)}{u^{2}+f(r)}=0. \nonumber \\ \label{g9}
\end{eqnarray}
By solving the above equations, we have
\begin{eqnarray}
\frac{d\ln u}{d\ln r}=\frac{D_{1}}{D_{2}}.\label{h1}
\end{eqnarray}
Also, we have
\begin{eqnarray}
D_{1}=\frac{r f'(r)}{2(u^{2}+f(r))}-2V^{2},\label{h2}
\end{eqnarray}
and
\begin{eqnarray}
D_{2}=V^{2}-\frac{u^{2}}{u^{2}+f(r)}.\label{h3}
\end{eqnarray}
From Eqs.(\ref{h1})-(\ref{h3}), we obtain the following result
\begin{eqnarray}
V^{2}=\frac{d\ln(p+\rho)}{d\ln \rho}-1.\label{h4}
\end{eqnarray}
For critical points, we take $D_{1}=D_{2}=0$ and get
\begin{eqnarray}
V_{c}^{2}=\frac{r f'(r)}{4f(r)+r f'(r)},\label{h5}
\end{eqnarray}
also
\begin{eqnarray}
u_{c}^{2}=\frac{1}{4} r f'(r),\label{h6}
\end{eqnarray}
where index $c$ stands for the value of the corresponding quantity at the critical point. The right side of Eq.(\ref{h4}) is necessarily positive. The range of the critical radius is determined by the following inequality
\begin{eqnarray}
4f(r)+r f'(r)>0.\label{h7}
\end{eqnarray}
By using Eq.(\ref{g2}), we have
\begin{eqnarray}
c_{s}^{2}=C_{4} \sqrt{[u^{2}+f(r)]^{-1}}-1,\label{h8}
\end{eqnarray}
which is the required equation for the speed of sound defined by $c_{s}^{2}=\frac{d p}{d\rho}$.

\section{Circular equatorial geodesics}

To analyze the circular motion of the test particle, we have to analyze the effective potential (Eq.(\ref{b5})) that is given by
\begin{eqnarray}
V_{eff}=\left(1-K \kappa N\left(\frac{N^{2}-1}{6}\right)-\frac{2m}{r}-\frac{\Lambda r^{2}}{3}\right)(1+\frac{L^2}{r^2}). \nonumber \\
\end{eqnarray}
In \textbf{Fig. 2a}, the graph of the effective potential of the massive particles versus $r$ is plotted. It expresses that the effective potential depends upon the angular momentum $L$. We observe the first extremum at $L=6$, while can not see any extremum for $L<6$. Also, the effective potential of the particles increases if the values of angular momentum $L$ increase. The solid circle in \textbf{Fig. 2a}, denotes the position of ISCO located at $r=6.089$. Moreover, effective potential $V_{eff}$ comprises two extrema for larger values of the angular momentum $L$, where the unstable circular orbit is located at the maximum of $V_{eff}$ and stable circular orbit located at the minimum of $V_{eff}$.

\begin{figure*}
a) \includegraphics[width=8.0cm]{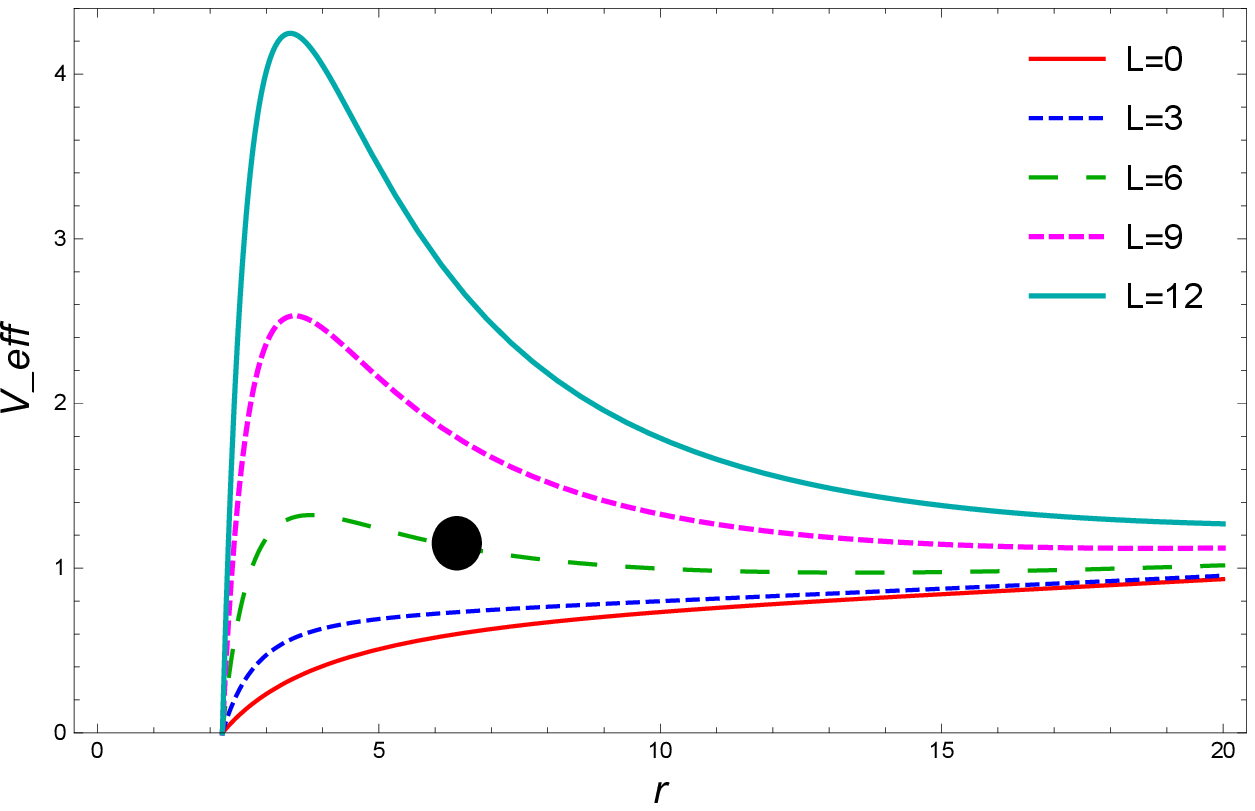}
b) \includegraphics[width=8.0cm]{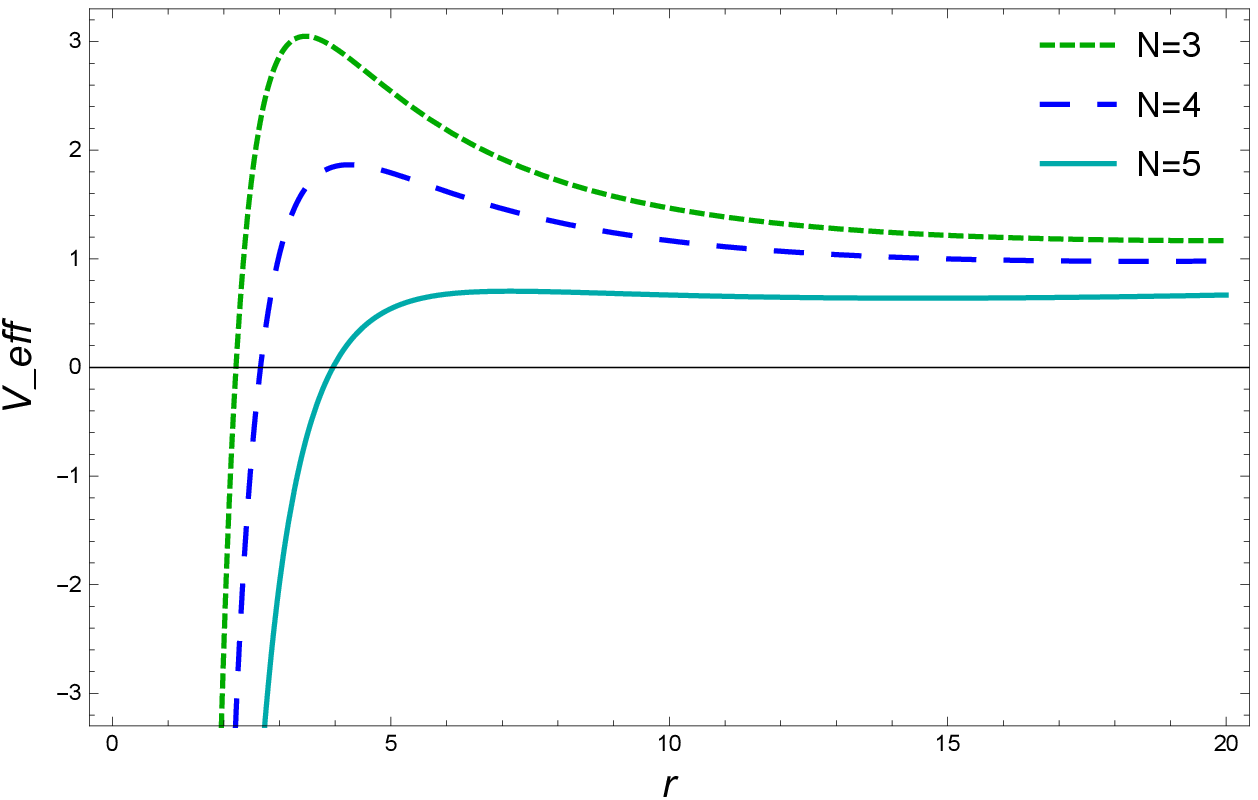}
\caption{ The graph illustrates the profile of $V_{eff}$ versus $r$  (a) for $N=3$, $\Lambda=-0.001$, and several values of $L$ (b) for $\Lambda=-0.001$, $L=10$ and different values of $N$}.
\end{figure*}

\textbf{Figure $2b$}, shows the profile of the effective potential $V_{eff}$ of the test particles and the dimensionless radius $r$. In plot $\textbf{2b}$, of effective potential $V_{eff}$, we take different values of parameter $N$ as $N=3$,$N=4$, and $N=5$ with fixed angular momentum $L=10$. Hence the effective potential of the test particles' motion decreases by increasing the values of parameter $N$.
The ISCO is the smallest marginally stable circular orbit in which a test particle can stably orbit around a massive object in GR.
The location of the ISCO depends upon the angular spin of the central object. The ISCO plays an important role in BH accretion discs since it marks the inner edge of the disc. Now we are interested to calculate ISCO of $SU(N)$ of non-linear sigma model BH, we are unable to find ISCO analytically, so we can find ISCO numerically by using the general formula of ISCO which is given in Ref. \cite{e1}. When $N = 3$, $K = 1$,  $\kappa=1/40$ and $\Lambda= -0.001$ the ISCO of $SU(N)$-NLSM BH is
$r_{isco} =6.089$, and further detail is given in \textbf{Table 1.}

To examine the accretion process around the BH,  the role of the ISCO is important but analysis of other radii is also compulsory. As it is studied earlier, the circular orbit exists whenever $r>r_{ph}$. For small perturbations, the particle's motion will be unstable if $r_{ph}<r<r_{isco}$. This relation implies that particles escape to infinity or fall into the BH. The motion of the particle remains in stable circular orbits if $r>r_{isco}$.
We can find out marginally bound orbit $r_{mb}$, photon sphere $r_{ph}$, and circular orbit $r_{circ}$ and by using Eqs. (\ref{c1})-(\ref{c3}), respectively which are given by

\begin{eqnarray}
r_{ph}=\frac{18m}{6-K \kappa N(N^{2}-1)}.
\end{eqnarray}

\begin{eqnarray}
r_{circ}>\frac{18m}{6-K \kappa N(N^{2}-1)}.
\end{eqnarray}
We are unable to find marginally bound orbit analytically, so we determine numerically which is given in \textbf{Table 1}.

In the equatorial plane, the following quantities can be calculated as
\begin{eqnarray}
E^{2}=\frac{(1-K \kappa N(\frac{N^{2}-1}{6})-\frac{2m}{r}-\frac{\Lambda}{3}r^{2})^{2}}{(1-K \kappa N(\frac{N^{2}-1}{6})-\frac{3m}{r})},
\end{eqnarray}
\begin{eqnarray}
L^{2}=\frac{r(m-\frac{\Lambda}{3}r^{3})}{(1-K \kappa N(\frac{N^{2}-1}{6})-\frac{3m}{r})},
\end{eqnarray}
\begin{eqnarray}
\Omega_{\phi}^{2}=\frac{m}{r^{3}}-\frac{\Lambda}{3},
\end{eqnarray}
\begin{eqnarray}
l^{2}=\frac{r(m-\frac{\Lambda}{3}r^{3})}{(1-K \kappa N(\frac{N^{2}-1}{6})-\frac{2m}{r}-\frac{\Lambda}{3}r^{2})^{2}}.
\end{eqnarray}
In \textbf{Fig 3}, we analyze the profile of specific angular momentum and the specific energy for parameter $N$ along with radius $r$ of the $ SU(N)$. In \textbf{Fig 3} \textbf{(a)}and \textbf{(b)}, we observe that as the value of $N$ increases the specific energy and specific angular momentum of the particle are also increased. Now we are interested to find the specific energy, the specific angular momentum, the angular velocity, and the angular momentum i.e ($E_{isco}$, $L_{isco}$, $\Omega_{isco}$, and $l_{isco}$ ) for innermost stable circular orbit but we can not determine all of these analytically so we calculate numerically which are given in the table.

\begin{figure*}
(a)\includegraphics[width=8.0cm]{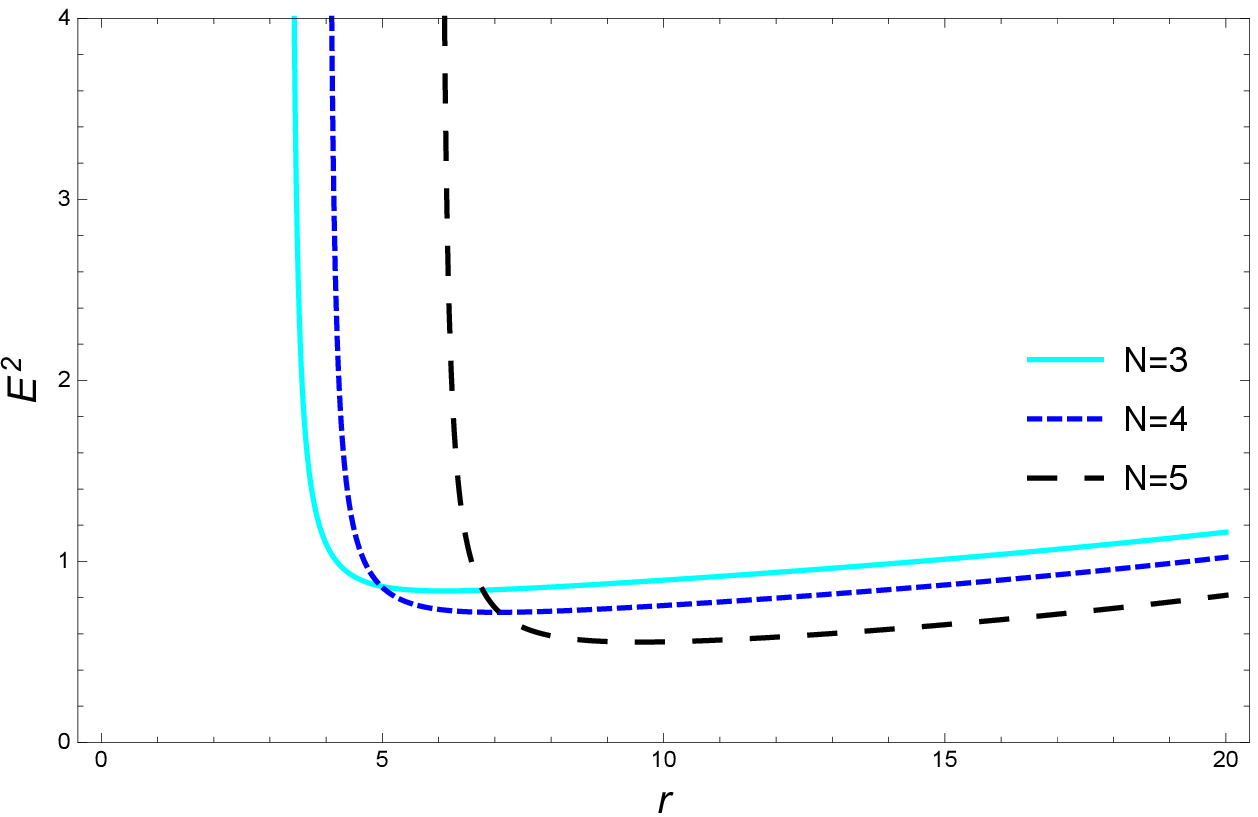}
(b)\includegraphics[width=8.0cm]{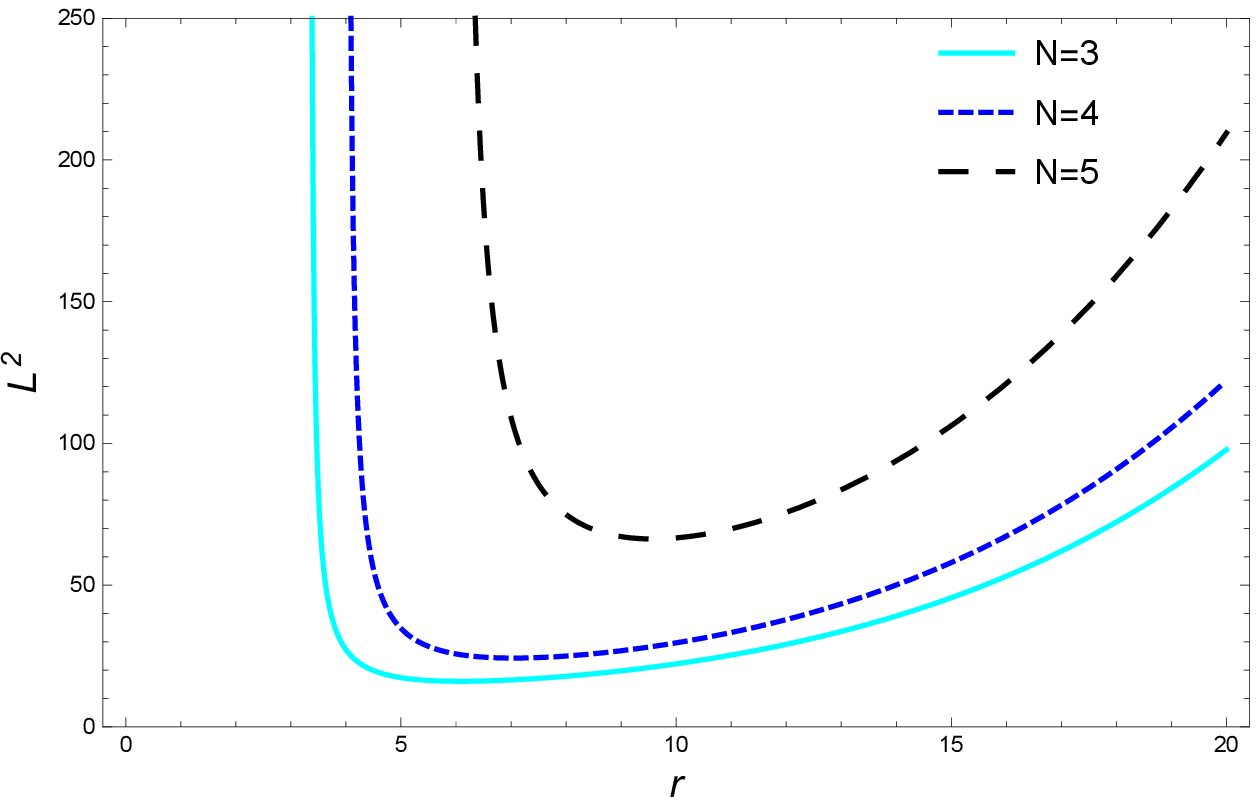}
\caption{ The profile of energy in (left plot) and angular momentum in (right plot) along radius $r$ is presented for various values of parameter $N$}.
\end{figure*}

\subsection{Mass evolution and critical points}
The radial velocity and the energy density of an isothermal fluid in the accretion disc of Einstein-$SU(N)$-NLSM BH are given by
\begin{eqnarray}
u=\frac{\sqrt{C_{4}^{2}-(k+1)^{2}(1-K \kappa N(\frac{N^{2}-1}{6})-\frac{2m}{r}-\frac{\Lambda}{3}r^{2})}}{(k+1)}.\nonumber\\
\end{eqnarray}
and
\begin{eqnarray}
\rho=\frac{C_{3}}{r^{2}} \frac{(k+1)}{\sqrt{C_{4}^{2}-(k+1)^{2}(1-K \kappa N (\frac{N^{2}-1}{6})-\frac{2m}{r}-\frac{\Lambda}{3}r^{2})}}. \nonumber\\
\end{eqnarray}
From Eqs.(\ref{g2}), (\ref{h5}) and (\ref{h6}), we get
\begin{widetext}
\begin{multline}
r_{c}=\Big(6 (k+1)^2 \Lambda \Big(\sqrt{6} \sqrt{(k+1)^6 \Lambda ^3 ((6 C^{2}_{4}+(k+1)^2(\kappa  K N(N^2-1)-6))^3+1944 (k+1)^6 \Lambda  m^2)}-\\ 108 (k+1)^6 \Lambda ^2 m\Big)^{1/3}\Big)^{-1}\\ \Big(\sqrt[3]{6}(\sqrt{6} \sqrt{(k+1)^6 \Lambda ^3((6 C^{2}_{4}+(k+1)^2(\kappa  K N(N^2-1)-6))^3+1944 (k+1)^6 \Lambda  m^2)}-\\ 108 (k+1)^6 \Lambda ^2 m)^{2/3}-6^{2/3} (k+1)^2 \Lambda (6 C^{2}_{4}+(k+1)^2(\kappa K N(N^2-1)-6))\Big),
\end{multline}
\end{widetext}
\begin{eqnarray}
V_{c}^{2}=\frac{-3m+r^{3}\Lambda}{9m-6r(1-K \kappa N(\frac{N^2-1}{6})-\frac{\Lambda}{2}r^{2})}.
\end{eqnarray}
and
\begin{eqnarray}
u_{c}^{2}=\frac{m}{2r}-\frac{\Lambda}{6}r^{2}.
\end{eqnarray}

\begin{figure}
\includegraphics[width=8.0cm]{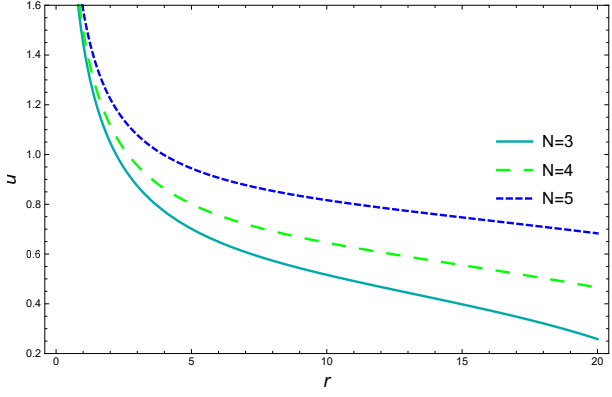}
\caption{The graph between $u$ and $r$ for certain values of $N$  and  fixed values of $k$ and $C_{4}$ are plotted}
\end{figure}

\textbf{Figure 4} depicts the radial velocity vs radius.
The fluid has zero radial velocity away from the BH and flows with subsonic speed prior to the critical points. After passing this point, in the neighborhood of the BH, the flow accelerates and enters the supersonic domain because of BH gravity. The radial velocity increases as the Einstein-$SU(N)$-NLSM BH parameter is $N$ increased, as shown in\textbf{ Fig.4}. 
\begin{figure}
\includegraphics[width=8.0cm]{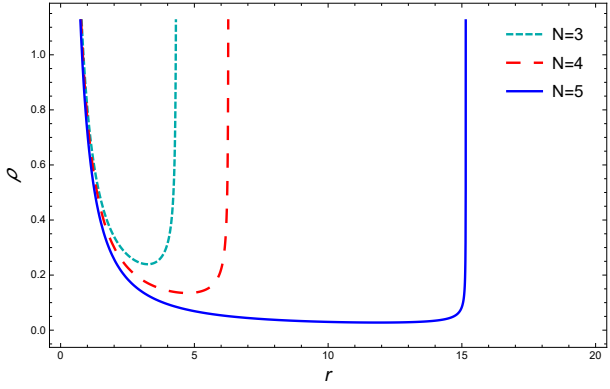}
\caption{ The profile for density of the fluid $\rho$ and radial distance $r$ is illustrated}.
\end{figure}
The \textbf{Fig. 5}, demonstrates the density curve of the fluid surrounding the BH for various values of $N$. The density increases when the value of parameter $N$ is increased. Furthermore, for isothermal fluids, the mass of the BH varies over time according to the following relation
\begin{eqnarray}
\dot{M}=4 \pi C_{1}(p+\rho) \sqrt{1-K  \kappa N(\frac{N^2-1}{6})-\frac{\Lambda}{3}r^2}   M^2, \nonumber\\
\end{eqnarray}
which shows that the mass accretion rate for Einstein-$SU(N)$-NLSM BH is quite different from Schwarzschild BH. The accretion rate entirely relies on the metric parameter $N$ and the characteristics of accreting fluid onto the Einstein-$SU(N)$-NLSM BH. If $\dot{M}$ is positive, so $(p+\rho)$ must also be positive.

\begin{figure}
\includegraphics[width=8.0cm]{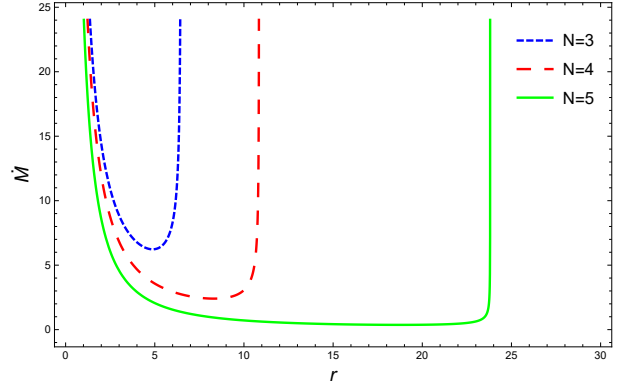}
\caption{ The profile shows the graph for accretion rate along dimensionless radius $r$ for various quantities of $N$}.
\end{figure}

\textbf{Figure 6} illustrates the graph of mass accretion rate versus $r$. For altered values of the parameter, $N$ the accretion rate decreases (increases) for smaller (larger) radii. It is observed that the accretion rate raised nearby the BH as a result of immense gravitational influence.

We obtained critical accretion time and mass by using Eq. (\ref{g7}), which are given by
\begin{eqnarray}
t_{cr}=[4\pi C_{1}(p+\rho)\sqrt{\Big(1-K \kappa N(\frac{N^2-1}{6})-\frac{\Lambda}{3} r^{2}\Big)}M_{i}]^{-1} \nonumber \\
\end{eqnarray}
and
\begin{eqnarray}
M_{t}&=&M_{i}\Big[1- 4\pi C_{1}(p+\rho) \nonumber\\
&&~~~~~~~~ \times \sqrt{\Big(1-K \kappa N(\frac{N^2-1}{6})-\frac{\Lambda}{3} r^{2}\Big)} M_{i}t \Big]^{-1}. \nonumber\\
\end{eqnarray}
We observe that for ordinary flow, the mass of BH rises with the increase of parameter $N$. In Ref. \cite{h1} Rodrigues et al. investigated the characteristics of a Schwarzschild BH in the context of a non-minimally coupled scalar field. They discovered that for BHs with initial masses less than a specific critical value, scalar field accretion can lead to mass decrease even in the absence of phantom energy and Hawking radiations.

\section{Radiant energy flux}

In the equatorial plane, we analyze the radiant flux from the outer layer of the accretion disc by using the values of $E$, $L$, and $\Omega_{\phi}$. Radiation energy flux for accretion disc is determined (by Eqs. (\ref{c7}) and (\ref{c8})) as follows
\begin{eqnarray}
K(r)&=&\frac{3\dot{M}}{8}\frac{m}{\sqrt{r(m-\frac{\Lambda}{3}r^{3})}} \left[\frac{3m}{r}-\left(1-K \kappa N \left(\frac{N^2-1}{6}\right) \right) \right] \nonumber \\
&& \times \left[\pi r^{4}(1-K \kappa N\left(\frac{N^2-1}{6}\right)-\frac{m}{r}-\frac{2}{3}\Lambda r^{2})^2\right]^{-1} \nonumber \\
&& \times\int^{r}_{r_{isco}}F(r)dr,
\end{eqnarray}
where
\begin{eqnarray}
F(r)&=&\frac{1}{2}\sqrt{\frac{m}{r}-\frac{\Lambda}{3}r^{2}} \nonumber \\
&&  \times \left[1-K \kappa N(\frac{N^2-1}{6})-\frac{m}{r}-\frac{2}{3}\Lambda r^{2}\right] \nonumber \\ 
&&\times \left[\frac{9m}{\Lambda r^{3}-3m}(1-K \kappa N(\frac{N^2-1}{6})-\frac{2m}{r}-\frac{\Lambda}{3}r^{2}) \right.\nonumber \\
&&~~~~~~ \left.+(4-4 K \kappa N(\frac{N^2-1}{6})-\frac{12m}{r})\right] \nonumber \\
&&\times \left[1-K \kappa N(\frac{N^2-1}{6})-\frac{3m}{r}\right]^{-2}.
\end{eqnarray}
To examine the behavior of radiation flux of the accretion disc around the $SU(N)$-NLSM BH for various values of BH parameter $N$ is plotted in \textbf{Fig 7}. We observe that by increasing $N$, the radiation flux of the accretion disc decreases.

\begin{figure}
\includegraphics[width=8.0cm]{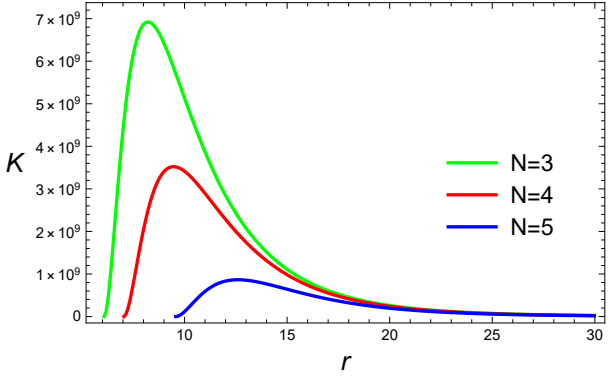}
\caption{The plot displays the energy flux $K$ verses $r$, for altered value of $N$}.
\end{figure}

\subsection{Radiation temperature}
Assuming that the accretion disc must be in thermal equilibrium, one could conclude that disc emission is black body radiation, which is described by the Stefan-Boltzmann law, which relates energy flux and temperature as
\begin{eqnarray}
K(r)=\sigma T^{4}.
\end{eqnarray}
Here $\sigma $ is the Stefan-Boltzmann constant. We analyze the temperature of the disc by using the parameter $N$. In \textbf{Fig.8}, the profile of disc temperature is drawn for various values of $N$ with $\Lambda=-0.001$. We examine that the temperature of the disc decreases as the value of the parameter $N$ increases.
\subsection{Radiative efficiency}
Radiation from gravitational energy is produced as the disc's material progressively spirals toward the center. Radiative efficiency, which is the ability of the central object to transform rest mass into radiation, is a quantity that can be determined by identifying the specific energy in the ISCO radius. The formula for radiative efficiency is given by
\begin{eqnarray}
\eta^{*}=1- E_{isco}
\end{eqnarray}
In \textbf{Table 1}, we display the numerical result of ISCO, marginally bound orbit $r_{mb}$, specific energy, specific angular momentum, specific angular velocity, angular momentum (i.e., $E^{2}_{isco}$, $L^{2}_{isco}$, $\Omega^{2}_{isco}$,$l^{2}_{isco}$), maximum energy flux, and maximum temperature distribution. Also, numerical investigation of all these quantities is represented in \textbf{Table 2}.

\begin{figure}
\includegraphics[width=8.0cm]{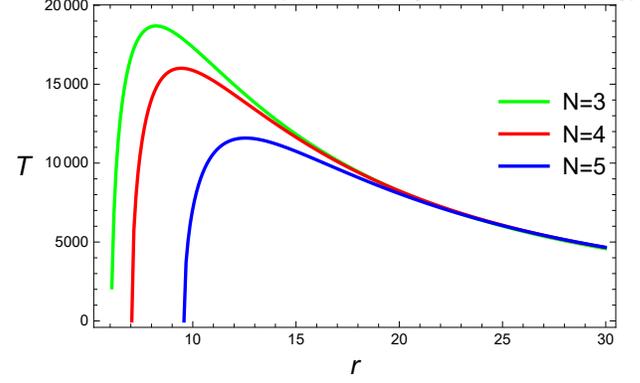}
\caption{The plot displays the relation between radiation temperature verses $r$, for altered value of $N$}.
\end{figure}

\begin{table*}
    \caption{The numerical values of ISCO, marginally bound orbit $r_{mb}$, specific energy, specific angular momentum, specific angular velocity, angular momentum for an innermost stable circular orbit, maximum energy flux, and maximum temperature distribution for $K = 1$,$\kappa=\frac{1}{40}$ and $\Lambda = -0.001$.} 
    \centering 
        \begin{ruledtabular}
    \begin{tabular}{ccccccccccc} 
        $N$ &$r_{mb}$ & ISCO \ &$ E^{2}_{isco}$ \ & $L^{2}_{isco}$ \ & $\Omega^{2}_{isco}$ \ & $l^{2}_{isco}$ & $K_{max}(r)$ & $T_{max}(r)$ &$\eta^{*}$\\ [0.7ex] 
        \hline 
              3&  (4.18471, 14.5354) &6.089&     0.83705 &         16.0743 &            0.00476 &        19.204 &         $6.88\times10^{9}$ &   $1.86\times10^{4}$ &0.16295 \\ 
       4  & (4.67857, 19.3415)& 7.046&     0.71863 &        24.2657 &            0.00319&         33.767 &         $3.5\times10^{9}$ &     $1.5980\times10^{4}$ & 0.28137\\
              5  &   (6.58478, 24.242) & 9.574&     0.55430 &       66.298 &            0.00147&          119.606 &         $8.59999\times10^{8}$ &   $1.1300\times10^{4}$ &0.4457 \\
    \end{tabular}
        \end{ruledtabular}
    \label{table:nonlin}
\end{table*}
\begin{table*}
    \caption{The numerical values of ISCO, marginally bound orbit $r_{mb}$, specific energy, specific angular momentum, specific angular velocity, angular momentum for an innermost stable circular orbit, maximum energy flux, and maximum temperature distribution for Schwarzschild case by considering $K = 0$, and $\Lambda = 0$.} 
    \begin{ruledtabular}
    \centering 
    \begin{tabular}{c c c c c c c c c} 
        ISCO &$ r_{mb} $ \ &$ E^{2}_{isco}$ \ & $L^{2}_{isco}$ \ & $\Omega^{2}_{isco}$ \ & $l^{2}_{isco}$ & $K_{max}(r)$ & $T_{max}(r)$ &$\eta^{*}$\\ [0.7ex] 
        \hline 
         6&4&     0.888 &         12 &            0.00463 &        13.5 &         $4.2\times10^{9}$ &   $1.6\times10^{4}$ & 0.112  \\ 
    \end{tabular}
    \end{ruledtabular}
    \label{table:nonlin}
\end{table*} 

\subsection{Epicyclic frequencies}
A particle will undergo small oscillations in both radial and vertical directions if the particle's motion on a circular orbit is affected by perturbations in the equatorial plane. We determine
radial and vertical epicyclic frequencies by using Eqs. (\ref{d6}) and (\ref{d7}), as follows
\begin{widetext}
\begin{eqnarray}
\Omega_{r}^2=\frac{-18m^{2}-4r^{4}\Lambda(1-K \kappa N(\frac{N^{2}-1}{6}))+3mr(1-K \kappa N(\frac{N^{2}-1}{6})+5r^{2}\Lambda)}{3r^{4}},
\end{eqnarray}
\end{widetext}
and
\begin{eqnarray}
\Omega_{\theta}^2=\frac{m}{r^{3}}-\frac{\Lambda}{3}.
\end{eqnarray}

\begin{figure}
\includegraphics[width=8.0cm]{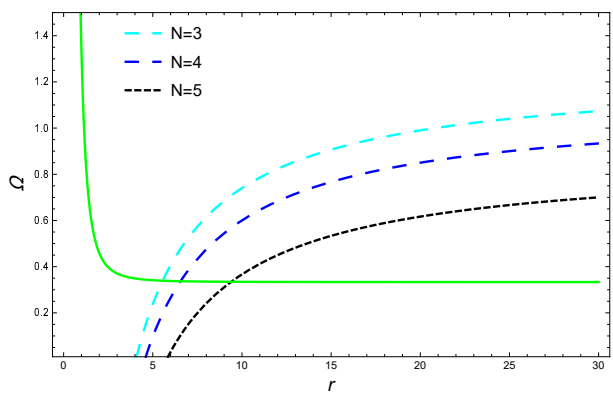}
\caption{The graphical representation of epicyclic frequencies and the green curve represents the behavior of vertical frequency}.
\end{figure}
In \textbf{ Fig. 9}, the graph of epicyclic frequencies versus the radial parameter $r$ is plotted. In the profile of epicyclic frequencies $(\Omega_{\theta}, \Omega_{r})$ along dimensionless radial parameter $r$, the green curve represents the behavior of vertical epicyclic frequency $\Omega_{\theta}$ which shows that the vertical frequency is independent on the parameter $N$ and completely rely on the radial distance $r$. It is observed from \textbf{Fig. 9} that the vertical frequency of the particles decreases by increasing the radius $r$. While dotted curves indicate the behavior for radial frequency. From these curves, it can be explained that the radial frequency of the particles decreases by increasing the values of the parameter $N$.
\begin{figure}
\includegraphics[width=8.0cm]{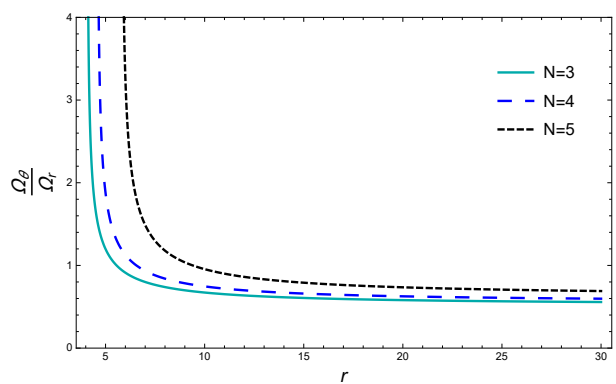}
\caption{ The plot illustrates the ratio of vertical and radial frequency}.
\end{figure}
From \textbf{Fig. 10}, we observe that the ratio of epicyclic frequencies is a decreasing function of $r$. It is also obvious that in the vicinity of BH, this ratio is very greater than unity but far from the BH it turns to unity and decreases by increasing $N$.

\section{Conclusions}

In this article, we examine the accretion and geodesic motion of particles around the Einstein-$SU(N)$-NLSM BH, in the equatorial plane. These orbits have been analyzed for their stability and circular geodesics, oscillations induced by small perturbations, unstable orbits, and ultimately the general formulation for the process of accretion flow in the vicinity of the BH has been presented. Also, the BH's effective potential, specific angular momentum, typical radius, specific energy, emission rate, epicyclic frequencies, dynamical parameters, and mass evolution are found. By assuming the equation of state $p = k\rho$ for an isothermal fluid, we can find some general solutions of fluid flow onto Einstein-$SU(N)$-NLSM BH. The effect of $SU(N)$-NLSM parameter $N$ is investigated for each aspect. The loci of the stable and unstable circular orbits are transformed to other locations under the $SU(N)$-NLSM parameter $N$ influence as per our observations. As the parameter $N$ increases, the effective potential $V_{eff}$ decreases, also we locate the position of the innermost stable circular orbit in \textbf{Fig. 2a}.  The positions of the characteristic radii such as $r_{isco}$, $r_{ph}$, $r_{sin}$, and $r_{mb}$ in this space-time deviate significantly from the Schwarzschild solutions. In \textbf{Table 1}, we analyzed the numerical results for ISCO, marginally bound, $E_{isco}$, $L_{isco}$, $\Omega_{isco}$, $l_{isco}$, and radioactive efficiency. In \textbf{Fig. 3}, the effect of parameter $N$ on energy and angular momentum of the BH is represented. Furthermore, as $N$ grows, the efficiency of accretion increases. As the BH parameter $N$ becomes larger, the radiation flux, radiant temperature, and energy efficiency of massive particles decrease. In addition to investigating circular orbits and their characteristics, epicyclic frequencies are investigated in this article. The vertical epicyclic frequency decreases monotonically with $r$ and has no extreme.

Finally, for the case of an isothermal fluid, the nature of the radial velocity, fluid particle density, and accretion processes are investigated using the equation of state parameter $k=0.6$. Apart from BH, fluids have zero radial velocity; hence, the radial velocity will be a decreasing function of $r$. When accretion occurs, fluid flows through a critical point where the flow rate and sound speed are equivalent. The fluid flows with subsonic speed before the critical point. The flow speed increases and then becomes supersonic after reaching that point and near the black hole due to the huge gravitational field. Finally, when the accretion rate is examined, it is found that it is affected by the metric parameter $N$ as well as the fluid's nature. In the case of normal fluid, the mass accretion rate increases because of the gravitational effect, and its value grows even more as one gets closer to the BH. In the case of the Schwarzschild black hole, the accretion rate also increases because of the positive deviation. It is essential to note that the particle being considered in this research is not spinning. An important factor in the path of a particle is its spin. Additionally, for the sake of clarity, the viscosity and the magnetic field of the accretion disc are disregarded and the fluid is considered to be ideal. However, these variables can affect the velocity of the test particle and extension, the accretion disc's structure, and the emission rate. To that end, we plan to look into how spinning particles and accreting viscous fluids react to the Einstein-$SU(N)$-NLSM BH under a magnetic field. {In the future it would interesting to construct the BH solution in the  Einstein- SU(N)-Skyrme model to see the effects of the Skyrme term explicitly on the accretion and accretion disk.}

\section*{Acknowledgements}

Tao Zhu is supported in part by the Zhejiang Provincial Natural Science Foundation of China under Grant No. LR21A050001 and LY20A050002, the National Key Research and Development Program of China under Grant No.2020YFC2201503, the National Natural Science Foundation of China under Grant No. 12275238, No. 11975203, No. 11675143, and the Fundamental Research Funds for the Provincial Universities of Zhejiang in China under Grant No. RF-A2019015.


\end{document}